\documentclass[preprint]{rmaa}
\usepackage{paralist}
\usepackage{psfrag,color}

\title{Optical Photometric and Spectroscopic study of the Seyfert Galaxy SBS 0748+499}
\author{J. Torrealba,\altaffilmark{1} E. Ben\'{\i}tez,\altaffilmark{1}
A. Franco-Balderas\altaffilmark{1} and V. H.
Chavushyan\altaffilmark{1,2}} \altaffiltext{1}{Instituto de
Astronom\'{\i}a, Universidad Nacional Aut\'onoma de M\'exico.}
\altaffiltext{2}{Instituto Nacional de Astrof\'{\i}sica, Optica y
Electr\'onica.} \shortauthor{Torrealba et al.} \shorttitle{study
of the Seyfert Galaxy SBS 0748+499} \fulladdresses{ \item Erika
Ben\'{\i}tez, Alfredo Franco-Balderas and Janet Torrealba:
Instituto de Astronom\'{\i}a, UNAM, Apartado Postal 70-264, 04510
  M\'exico D.F., M\'exico (\email{erika;alfred;cjanet@astroscu.unam.mx}).

\item V. H. Chavushyan: Instituto Nacional de Astrof\'{\i}sica,
Optica y Electr\'onica, Apartado Postal 51 y 216, 72000 Puebla,
Pue., M\'exico (\email{vahram@astroscu.unam.mx,
vahram@inaoep.mx}).}

\listofauthors{J. Torrealba, E. Ben\'{\i}tez, A. Franco-Balderas
\& V. H. Chavushyan} \indexauthor{Torrealba, J.}
\indexauthor{Ben\'{\i}tez, E.} \indexauthor{Franco-Balderas, A.}
\indexauthor{Chavushyan, V. H.}

\abstract{We present the first optical photometric study of the
active galaxy {\sc{SBS}} 0748+499. First, we present $B$,
$V$, $R$ and $I$ photometric data: total magnitudes and $B-V$,
$B-R$, $B-I$ colors; surface brightness, color and geometric
profiles, with emphasis on the morphology and its relation to the
global photometric properties of this galaxy. Then, from our
surface photometry study we derive the bulge--to--disk
luminosity ratio $B/D$ in the four bands. We find that the host
galaxy shows a bar ($a \sim 8$ kpc) and a low--brightness spiral
structure. The morphological classification for the host galaxy of
this AGN is SBab. Additionally, we present new optical
spectrophotometric observations that clearly show that the object
can be classified as Sy 1.9 galaxy. Finally, we estimate the
black hole mass ($M_{BH}$) associated with the nucleus of
SBS~0748+499 using the absolute $R$--band bulge
magnitude--$M_{BH}$ relation, and the FWHM
[OIII]$_{5007}$ -- $\sigma_{\star}$ relation. We find that
$M_{BH}= 2.6\,\times\,10^7\,M_{\odot}$ and $M_{BH}=
8.8\,\times\,10^7\,M_{\odot}$, respectively.}

\addkeyword{Galaxies: Active Galaxies} \addkeyword{individual
(SBS~0748+499)} \addkeyword{Galaxies: Surface Photometry}
\addkeyword{Galaxies: Seyfert}

\begin{document}

\maketitle

\section{Introduction}

Since the work of \citet{KW74} Seyfert galaxies were divided into type
1 (Sy1) and type 2 (Sy2) active galactic nuclei (AGN).  This
classification is based on their optical spectra: Sy1 are those
Seyferts that show broad permitted lines with FWHM greater than $\sim$
10$^{3}$ km s$^{-1}$ and also narrow forbidden lines with FWHM $\sim$
10$^{2}$ km s$^{-1}$ while Sy2 show permitted and forbidden narrow
lines. Later, \citet{Os81} introduced intermediate Seyfert objects
that run from 1.2 to 1.9. In particular, Sy 1.8 show weak, but readily
visible broad H$\alpha$ and H$\beta$ emission, while Sy 1.9 show only
broad H$\alpha$. A new class of Seyfert galaxies showing relatively
narrow permitted lines but with line ratios and FeII emission similar
to the Sy1, was introduced by \citet{OP85}. Objects in this class are
known as ``narrow line Seyfert one'' (NLS1).

Besides their optical spectral characterization, it is also well known
that Seyfert galaxies are nearby active galaxies with hosts galaxies
showing large-scale morphologies that resemble those of normal or
inactive galaxies \citep[see][]{Ta96,M99,B01}.  On the other hand,
several studies have concluded that the black holes masses are
directly proportional to some properties of the galaxy bulge
component, e. g. to the mean velocity dispersion of the bulge, the
luminosity of the bulge component, etc. We refer the reader to
\citet{KG01} for a review on this subject. It is also known that only
a handful of measurements of black hole masses in Seyfert galaxies
have been obtained using essentially reverberation mapping techniques
\citep[see][]{W99,K00,W02}. However, this method is limited only to
Sy1 since the broad line region of Sy2 is hidden from our view,
according to the unified model. Recently, \citet{B03} showed that a
real correlation exists between the FWHM [OIII] and $\sigma_{\star}$
in QSOs and Sy1 galaxies, although with a large scatter. Nevertheless
it can predict the black hole mass ($M_{BH}$) to within a
factor of 5.

Recently, we have studied the multiwavelength properties of a sample
of 25 NLS1 isolated from the Second Byurakan Survey \citep[SBS,
see][]{S03}. Measuring the line emission ratios of this sample, we
discovered that SBS~0748+499 does not show evidences of FeII
emission. It was nevertheless classified as NLS1 by \citet{Ste02},
although this is clearly a different type of Seyfert. We also noticed
that the dimensions of this object ($\sim$\,25 $\arcsec$ $\times$ 25
$\arcsec$) allow us to study the host galaxy associated to its nucleus
with the 1.5m telescope at OAN--SPM\footnote{Observatorio
Astron\'omico Nacional, San Pedro M\'artir, Baja California,
M\'exico.}. Therefore, we decided to study its optical photometric and
spectroscopic properties in a separate work. Thus, one of the goals of
this paper is to establish its fundamental properties using surface
photometry techniques, and also to obtain a more accurate
spectroscopic classification of this object using new high
signal to noise ratio S/N and moderate resolution
spectroscopic observations. A second goal will be the application of
the correlation FWHM [OIII]--$\sigma_{\star}$ and to estimate the
$M_{BH}$ associated to SBS~0748+499, and then compare this result with
the one obtained with a different method \citep[see][]{MD01}, based on
our estimation of the absolute $R$-band bulge magnitude $M_{R}$ for
this object.

Finally, we want to point out that the $M_{BH}$ is a fundamental
property of AGN because it governs the physics of the central
engine. More measurements of black hole masses in nearby galaxies
are needed, over the widest possible range of host galaxy types
and velocity dispersions, in order to obtain a definitive
present--day black hole census \citep[see][]{B04}. Having this in
mind, and considering the fact that we obtain the structural
parameters of SBS~0748+499 and its morphological classification,
along with its optical spectra, we will be able to give an
estimation of $M_{BH}$ that we assume inhabits the nucleus of this
object, using two different methodologies, under the assumption
that they can be applied to intermediate Seyfert galaxies.

The structure of the paper is as follows. The details of the
observations and data reduction are given in \S~\ref{sec:Obs}. We show
the results of the aperture photometry in \S~\ref{sec:mag}.  In
section \S~\ref{sec:surface}, the geometric and surface brightness
profiles of SBS~0748+499 are shown and we present the parameters
obtained with the analysis of the profile decomposition that is used
to give a morphological classification for the host galaxy. Also,
based on our spectroscopical analysis we give the spectral
classification of the galaxy in \S~\ref{sec:espectro}.  The parameters
obtained in the two later sections are used to give a rough estimation
of the black hole mass of the galaxy in \S~\ref{sec:BH}. Finally, in
section \S~\ref{sec:discusion} we present a discussion and a summary
of the results obtained in this work.

\section{Observations}
\label{sec:Obs}
\begin{figure}[!t]
  \includegraphics[width=\columnwidth]{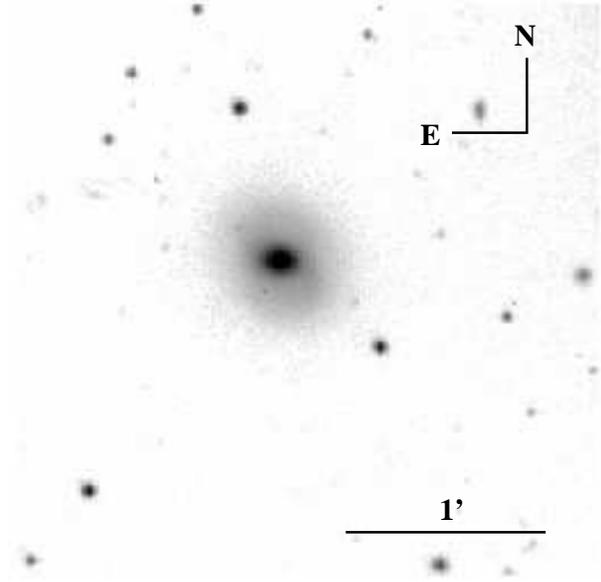}
  \caption{$B$--band image of SBS~0748+499, barely showing the presence a low-brightness
pseudo-spiral structure. The cosmology used in this paper is
$H_0$=75\,$\mathrm{km\,s^{-1}\,Mpc^{-1}}$, therefore, 1\arcsec=0.47
kpc.} \label{fig:RGB}
\end{figure}

The coordinates of the Seyfert galaxy SBS~0748+499 are:
$\alpha$\,=\,07$^{\mathrm{h}}\,$51$^{\mathrm{m}}\,$51.9$^{\mathrm{s}}$
and $\delta$\,=\, $+$49\,$^{\circ}\,$48$\arcmin\,$51$\farcs6$
(J2000). Based on low-resolution optical spectroscopy, its
redshift is $z=0.0244$.

We have obtained broad band $BVRI$ images of SBS~0748+499 in the
Johnson-Cousins system with the 1.5m telescope at the OAN-SPM. The
observations were carried out on november 24$^{th}$ (UT) 2001. We
have used a 1024\,$\times$\,1024 CCD with a binning of
$2\,\times\,2$ pixels, that gives a scale of plate $\sim
0\farcs51$/pix. With this configuration we covered an area of
$4\farcm3\times 4\farcm3$ in the sky. The integration times were
2400, 1200, 600 and 400 seconds for {\it{B}}, {\it{V}}, {\it{R}}
and {\it{I}}, respectively. We have determined an average seeing
of $\sim\,2\,\arcsec$ in the four bands using six field stars
present in the images. Images in the four filters are uniform,
with a sky-brightness of $V=21.50\pm0.01$.

Standard IRAF\footnote{~IRAF is the Image Reduction and Analysis
Facility distributed by the National Optical Astronomy
Observatories, which is operated by  the Association of
Universities for Research in Astronomy (AURA) under agreement with
the National Science Foundation (NSF).} procedures for data
reduction were used, i.e., bias and dark subtraction,
flat-fielding, cosmic-ray removal and sky subtraction to produce
the final images. The images were flat--fielded using sky flats
taken in each filter at the beginning of the night.
Figure~\ref{fig:RGB} shows the $B$--band image of SBS~0748+499.

The photometric calibration was done using three standard stars
selected from Landolt's list \citep[]{L92}. The equations to transform
instrumental magnitudes into standard magnitudes were taken from
\citet{FrBa03}. We observed no significant deviations between our
magnitudes obtained and the standard star magnitudes.  Therefore, the
errors in the calibration (quadratic sum of the individual errors)
are: $\sigma_{B}\,\sim\,0.009$, $\sigma_{V}\,\sim\,0.006$,
$\sigma_{R}\,\sim\,0.008$ and $\sigma_{I}\, \sim\,0.009$.

\begin{figure}[!t]
\vskip 7mm
  \includegraphics[width=\columnwidth]{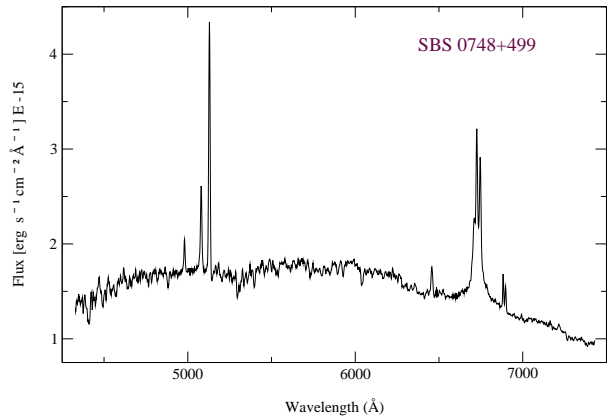}
  \caption{SBS~0748+499 spectra obtained with GHO 2.1m telescope,
Cananea, Sonora.}
  \label{fig:espectro}
\end{figure}

Additionally, we have obtained intermediate resolution spectra of
SBS~0748+499 with the INAOE's 2.1m telescope at the
GHO\footnote{Guillermo Haro Observatory  at Cananea, Sonora,
M\'exico} in march of 2002, see Figure~\ref{fig:espectro}. The
Boller \& Chivens spectrograph was set in the 4000 -- 5700 \AA\
and 5500 -- 7200 \AA\ ranges with a resolution 1.2\,\AA/pix,
combined with the deployed slit width $2\farcs5$, that gives an
overall resolution of 6 \AA\ FWHM. The S/N ratio was approximately
40 in the continuum near H${\beta}$ and H${\alpha}$. The spectra
were reduced using IRAF routines for long-slit spectroscopy. We
observed also standard stars for flux calibration.

\section{Photometry}
\label{sec:mag}

We have computed apparent and absolute total magnitudes in
$\it{BVRI}$ bands in two different ways. First, we have used a
polygonal aperture chosen interactively with IRAF/APPHOT/POLYPHOT
in a way that it contains the entire galaxy (radius $\sim\,60\,\arcsec$). 
Second, with the IRAF/ISOPHOTE/ELLIPSE, we
used elliptical aperture until $\mu_{B}=\,$25 mag arcsec$^{-2}$
was reached, which corresponds to a semi-major axis lenght of
$19\farcs7$ ($9\,\mathrm{kpc}$), and also up to $50\farcs6$
(24$\,\mathrm{kpc}$) in order to compare them later with the
results obtained with the polygonal photometry.

Apparent magnitudes were corrected for atmospheric
extinction using the values that we obtained for the OAN--SPM:
0.235$\,\pm\,$0.010, 0.147$\,\pm\,$0.005, 0.079$\,\pm\,$0.004 and
0.053$\,\pm\,$0.005 for the $\it{B, V, R}$ and $\it{I}$ filters,
respectively. These values are in agreement with those reported
by \citet{Sch82}. Then, these magnitudes were transformed to the
standard system. For absolute magnitudes, the correction for
Galactic extinction was made using the extinction data presented
in NED\footnote{NASA/IPAC Extragalactic Database.} from the dust
extinction maps of \citet{SFD98} based on IRAS/DIRBE measurements
of diffuse IR emission. We also applied K--corrections
interpolated from \citet{F&G94} and inclination corrections from
\citet{Tully98} using the coefficients given by \citet{Ver01}. The
corrections mentioned above were 0.51, 0.40, 0.34 and 0.26 mag for
the $\it{B}$, $\it{V}$, $\it{R}$ and $\it{I}$ filters,
respectively.

Calculations throughout this paper were done assuming that
$H_0$=75\,$\mathrm{km\,s^{-1}\,Mpc^{-1}}$. We have used z=0.0244---
obtained from our spectroscopic data, this also agrees with the value
reported by \citet{Ste02}--- and estimated a distance of 98\,Mpc for
SBS~0748+499. In Table~\ref{tab:mag} we present the apparent and
absolute total magnitudes obtained with the two procedures mentioned
above. The estimated accuracies in the given apparent magnitudes are
$\pm\,0.04\,\mathrm{mag}$ in $\it{B}$, $\pm\,0.01\,\mathrm{mag}$ in
$\it{V}$, $\pm\,0.02\,\mathrm{mag}$ in $\it{R}$ and $\it{I}$
filters. The magnitudes obtained for SBS~0748+499 with polygonal and
elliptical apertures at 24$\,\mathrm{kpc}$ are in very good agreement
in all bands, except for the $\it{I}$--band, where a small difference
of 0.15$\,\mathrm{mag}$ was found.

There is only one optical photometric data point
for SBS~0748+499 in the literature given by \citet{ZH66}.
They found a value of $B=15.1$ inside an aperture with 18\arcsec of radius.
This is consistent with our determination of $B=15.2$,
that was obtained using an elliptical aperture of
$a=19$\arcsec or $9\,\mathrm{kpc}$. Through polygonal and elliptical
($9\,\mathrm{kpc}$) apertures we obtained that $M_{B}=-19.9$.
This value is in the range given by \citet{Bo02} for a sample of
Sy galaxies ($-18.7$ to $-22.3$) with an average value $M_{B}=-20.7$.
From the above estimate, it is clear that SBS~0748+499 is 1.2
$\,\mathrm{mag}$ fainter than the mean value
$M_{B}=-21.1$ reported for normal S0a--Sa UGC sample galaxies
\citep[]{RH94}.

Finally, the corrected total colors obtained up to $\mu_{B}=\,$25
mag arcsec$^{-2}$ are: $B-V=0.84$, $B-R=1.39$, $V-R=0.55$ and
$V-I=1.13$. For the elliptical aperture (24 kpc) we obtained
$B-V=0.85$, $B-R=1.41$, $V-R=0.55$ and $V-I=1.13$. For the
polygonal aperture the colors are the same except $V-I=1.24$.
These colors are in agreement with the reported average values for
Seyfert galaxies given by \citet{Chatz01}.

\begin{table*}
 \begin{center}
 \caption{Apparent and Absolute total magnitudes.}
 \label{tab:mag}
  \begin{tabular}{ccccccccc}
\hline\hline
    & \multicolumn{8}{c}{Aperture}\\
    \cline{2-9}\\
 & \multicolumn{2}{c}{Polygonal}&
 &\multicolumn{2}{c}{Elliptical} &
 &\multicolumn{2}{c}{Elliptical}\\

& \multicolumn{2}{c}{($60\farcs0$)}&
 &\multicolumn{2}{c}{($19\farcs7$)} &
 &\multicolumn{2}{c}{($50\farcs6$)}\\

    \cline{2-3}\cline{5-6}\cline{8-9}\\
Filter &  m & M  &  & m & M  &  & m & M \\

\hline\noalign{\smallskip}
$B$ & $15.55$ & $-19.90$ && $15.73$ & $-19.71$ && $15.56$ & $-19.89$\\
$V$ & $14.58$ & $-20.75$ && $14.77$ & $-20.56$ && $14.58$ & $-20.75$\\
$R$ & $13.96$ & $-21.31$ && $14.16$ & $-21.10$ && $13.97$ & $-21.30$\\
$I$ & $13.20$ & $-21.99$ && $13.50$ & $-21.69$ && $13.35$ & $-21.85$\\
\hline\noalign{\smallskip}
\end{tabular}
\end{center}

Note.--- The elliptical aperture of $19\farcs7$ is characterized
by a surface brightness $\mu_{B}=25\,\mathrm{mag\,arcsec^{-2}}$ and
corresponds to a semi--major axis length of 9$\,\mathrm{kpc}$.
Elliptical aperture of $50\farcs6$ or 24$\,\mathrm{kpc}$ contains
almost the entire flux of the galaxy.

\end{table*}

\section{Surface Photometry}
\label{sec:surface}
\subsection{Geometric profiles}
\label{subsec:geometric}

In order to obtain the radial profiles and geometric parameters of
SBS~0748+499, we have used the IRAF/ISOPHOTE/ELLIPSE routine to fit
the elliptical contours. This routine is based on the iterative method
described by \citet{Jed87}. The contours are characterized mainly by:
(1) The semi-major axis length ($a$); (2) the mean intensity ($I_m$);
(3) the ellipticity $\epsilon=1-b/a$, where $b$ is the semi-minor
axis; (4) the position angle (PA); and (5) the parameter $a_4/a$. Note
that the $a_{4}/a$ parameter is a dimensionless measure. It is related
with the fourth cosine coefficient of the Fourier series expansion of
deviations from a pure ellipse, and gives information about the
isophotal shape.  So, if $a_{4}/a>0$ the isophote is {\it{disky}}
(pointed ends), if $a_{4}/a<0$ the isophote is {\it{boxy}} and if
$a_{4}/a=0$ then we have a perfect ellipse \citep[]{BM87}. For more
details see \citet{FrBa05}.

\begin{figure}[!t]
  \includegraphics[width=\columnwidth]{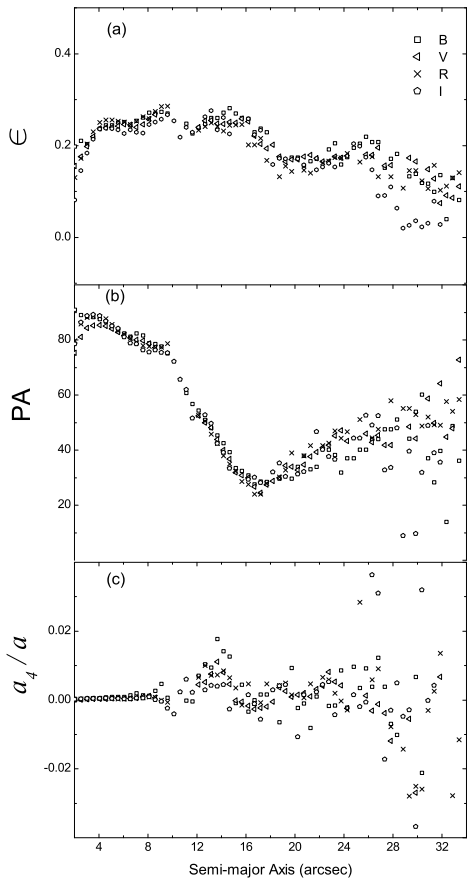}
  \caption{Geometric Profiles;
~(a)~Ellipticity, ~(b)~Position angle and ~(c)~Parameter {\it{a$_4$/a}}
vs semi-major axis {\it{a}}.}
  \label{fig:pargeo}
\end{figure}

To compute the geometric parameters we have run the routine in
free mode giving arbitrary initial values for $\epsilon$, PA and
$a$. The center of each fitted ellipse matched the center of
the galaxy obtained with IRAF/DAOFIND routine. Errors in the
intensity were obtained from the rms scatter of intensity data
along the fitted ellipse, while errors in $\epsilon$ and PA are
computed by ELLIPSE \citep[]{B96}. Figure~\ref{fig:pargeo} shows
that the behavior of $\epsilon$, PA and $a_4/a$ as a function of
$a$ is roughly the same in the four filters. Typical errors are
$\Delta\epsilon=0.03$, $\Delta$PA$=8\,^{\circ}$ and
$\Delta\,a_4/a=0.008$.

From the behavior of the geometric parameters we can distinguish
basically four different regions:

Region I (2\arcsec --4\arcsec): It is
dominated mainly by seeing effects, the isophotes are
quasi-circular while the PA has a nearly constant value of
$87\,^{\circ}$.

Region II (4\arcsec --10\arcsec): From 6\arcsec--10\arcsec~it
can be seen that $\epsilon$ grows 14$\%$ and reach the maximum value
in this region, PA diminishes to $11\,^{\circ}$, meanwhile
$a_{4}/a\sim\,0$. The smooth change in PA is considered as an
isophotal twist and could represent a triaxial system accordingly with
\citet{Wo95}. \citet{Elme96} performed a study of the isophotal twist
in barred galaxies and found that it is related with flat bars. These
kinds of bars are present only in early type galaxies (SBa--SBbc). They
also found that flat bars are characterized by a continuous increase
in the ellipticity. Therefore, this behavior could suggest the
existence of a barred-structure in this galaxy. Nevertheless, it is not
clearly seen in the brightness profiles.

Region III (10\arcsec --18\arcsec): In this region, we observe
that $\epsilon\sim\,0.24$ and that PA is systematically decreasing
towards $34\,^{\circ}$. Also, $a_{4}/a$ is maximum at 14\arcsec,
showing that the isophotes are $\it{disky}$. This led us to
suggest, accordingly with \citet{Elme96}, that this galaxy has
low-brightness spiral-arms embedded in the disk component. This
spiral structure is barely visible in our $B$-band image in Figure
1.

Region IV ($a > 18\arcsec$): The isophotes
becomes circular again fitting the disk's form. At the same time,
PA and $a_{4}/a$ shows larger dispersion due to the diminishing
of the intensity of the galaxy, which tends to be near the sky
brightness value.

\subsection{Surface Brightness Profile}
\label{subsec:brightness}

Surface brightness profiles were obtained by fitting ellipses with
a fixed position angle and ellipticity previously determined on
the external non-disturbed isophotes.  The characteristic values
for the fit were: $\epsilon=0.16\pm0.01$ and
PA$=34\,^{\circ}\pm1\,^{\circ}$. The maximum extension for the
brightness profiles was set up to 100 pixels or
$24\,\mathrm{kpc}$. The optical brightness of
the galaxy is totally contained within this radius. Once we have obtained the
azimuthal mean intensity for each fitted isophote, we transformed
it to instrumental surface brightness using the standard procedure
described in \citet{FrBa05}.

\begin{figure}[!t]
  \includegraphics[width=\columnwidth]{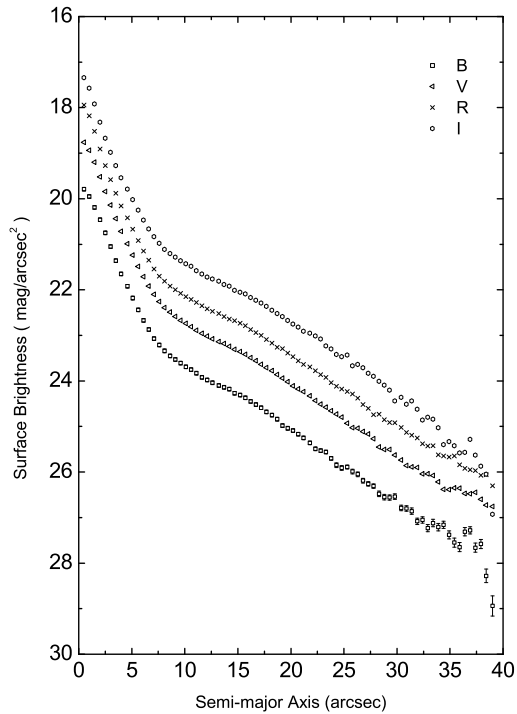}
  \caption{Surface brightness profiles of the galaxy
of SBS~0748+499 in {\it B},{\it V},{\it R} and {\it I}
filters. The value of
$\mu_{B}$=25\,mag\,arcsec$^{-2}$
was reached at $9\,\mathrm{kpc}$.
This isophote was characterized by $\epsilon=0.16\pm0.01$
and PA$=34\,^{\circ}\pm1\,^{\circ}$.}
  \label{fig:brillo}
\end{figure}

In Figure~\ref{fig:brillo} we show the surface brightness profile
for {\it{B}}, {\it{V}}, {\it{R}}, and {\it{I}} filters, all
corrected for atmospheric extinction and transformed to 
Landolt's photometric system. We have plotted the surface
brightness up to a radius of 39\arcsec (18.6\,kpc). This was done
because the errors are greater than 1$\%$ at this radius. Typical
errors in surface brightness were: $\Delta\mu_{B}=0.05$,
$\Delta\mu_{V}=0.02$, $\Delta\mu_{R}=0.03$ and
$\Delta\mu_{I}=0.03$. For clarity, we just show the errors in
{\it{B}} band. Also in Figure~\ref{fig:brillo} we observe that
$\mu_{B}=\,$25 mag\,arcsec$^{-2}$ was reached until a semi--major
axis length of $19\farcs7$, then the estimated semi--minor axis
length is $16\farcs5$, therefore the inclination angle respect to
the line of sight is {\it{i}}$\,\approx 33\,^{\circ}$
($i=\cos^{-1}\,b/a$) for SBS~0748+499.

\begin{figure}[!t]
  \includegraphics[width=\columnwidth]{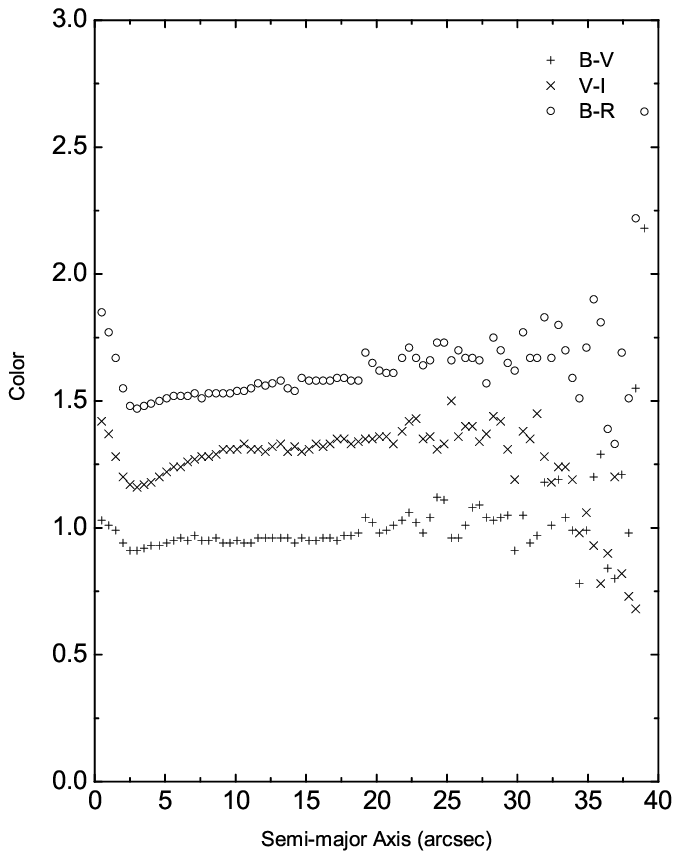}
  \caption{The behavior of the colors $(B-V)$, $(V-I)$ and $(B-R)$
of SBS~0748+499 is shown in this plot. It can clearly be observed
that the center is the reddest part, and that colors appear
flatter in the region between 2\arcsec--20\arcsec (1--9.5 kpc).
The dispersion is larger in the external parts of the galaxy due
to the increasing contribution of the sky.}

\label{fig:color}
\end{figure}

During the observations, we obtained the characteristic value
for the night sky in each band: $22.99\pm0.05, 21.50\pm0.01,
20.50\pm0.02$ and $18.68\pm0.04$ for {\it{B}}, {\it{V}}, {\it{R}} and
{\it{I}}, respectively.  Additionally, Figure~\ref{fig:color} shows
$B-V$, $B-R$ and $V-I$ color profiles which have been computed using
the values obtained from the surface brightness profile.

\subsubsection{Profile Decomposition}
\label{subsubsec:des}

The surface brightness profile analysis provides information about the
main components of the galaxy. The spiral galaxy bulges are in
fact not universally described with an $r^{1/4}$ law \citep[]{A95}.
Rather, a continuous range of bulge light profile shapes are now
known to exist (see Graham 2001 and references therein;
for Sy galaxies see Chatzichristou 2001).

The bulge observed in SBS~0748+499
was characterized by an effective radius ($r_e$) and its
corresponding surface brightness ($\mu_e$), a disk with scale
length ($h$) and its extrapolated central surface brightness
($\mu_0$). These four quantities are used to obtain the
bulge--to--disk luminosity ratio ($B/D$) that is useful to establish the
morphological classification for the host galaxy of SBS~0748+499.

In order to derive the physical parameters for the components of
the galaxy, we have modelled the surface brightness profile with an
exponential disk plus a S\'ersic $r^{1/n}$ profile \citep[]{S68}
with index {\it{n}}. The S\'ersic profile can be written as

\begin{equation}
\label{eq:one}
I(r)=I_e\exp{\Bigg\{-b_{n}\left[\left(\frac{r}{r_e}\right)^{1/n}-1\right]\Bigg\}}
\end{equation}
where the parameter $b_n$ is determined from the definition of
effective radius $r_e$ and $I_e$ is the surface brightness (or
half--light) at $r_e$ in intensity units or $\mu_e$ in mag
arcsec$^{-2}$ units. From \citet{MC2002} $b_n$ can be well
approximated by $b_{n}=1.997\,n-0.327$. When {\it{n}}\,=\,4 the
S\'ersic model reduces to the $r^{1/4}$ law; for $n=1$ reduces to
an exponential profile  $I(r)=I_{0}\exp{\{-r/h\}}$ , if
the central surface brightness of the disk is given by
$I_{0}=I_{e}e^{b_n}$ ($\mu_{0}$ in mag arcsec$^{-2}$ units) and
the disk scale lenght is $h=r_{e}/b_{n}^{n}$.

\begin{figure}[!t]
  \includegraphics[width=\columnwidth]{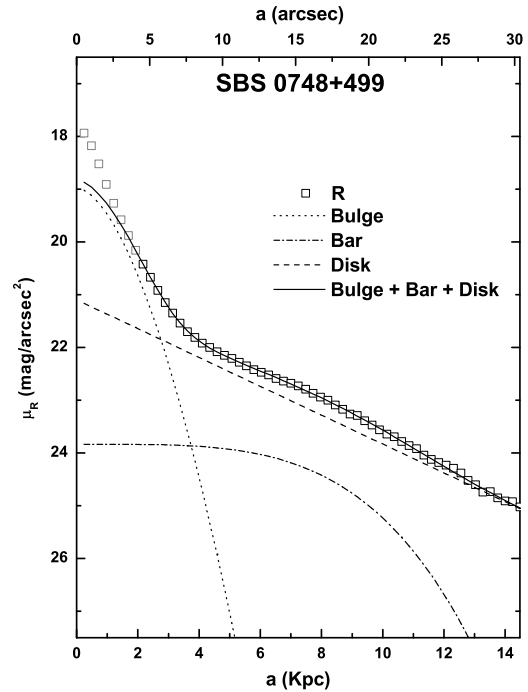}
  \caption{Profile decomposition for SBS~0748+499 in the {\it {R}}
band. The light profile of this galaxy is remarkably well
described by three components: a bulge and a bar modelled with a
S\'ersic profile and an exponential disk. In the center of the
profile the model differs from the observations because an
inner--cut off of 2\arcsec was applied during the decomposition
procedure.}
\label{fig:descomposicion}
\end{figure}

The decomposition profile was performed as follows. From
Figure~\ref{fig:brillo} we can clearly distinguish an inflection
point on the light profile that separates the region where the bulge
dominates over the disk. The bulge component is dominant up to
7$\arcsec$ and the disk component appears roughly linear up to 30
$\arcsec$. Then the profile becomes noisy, but the disk extends
over 40$\arcsec$. Once we have made the distinction between the
evident components in the profile, a simultaneous two--component
fit was applied until convergence was achieved. That is, at the
beginning, we iteratively fitted a function with five parameters:
three for the bulge component and, since $n=1$, two for the disk.
The $\chi^2$ statistic has been used as an estimator for the
quality of the fit. Then, for the geometric profiles and
minimization of $\chi^2$, we found that the best fit for the
profile was obtained taking into account an additional component:
a bar that can be modelled with a S\'ersic profile. The bar
component was characterized by a scale radius ($r_{bar}$) and its
corresponding surface brightness ($\mu_{bar}$). The profile fit
was performed using an inner cut--off radius of approximately
$2\arcsec$ to avoid the seeing effects. We have chosen for the
bulge component the region up to $7\farcs1$ (3$\,\mathrm{kpc}$);
for the disk we have worked from $7\farcs1$ up to $30\arcsec$
(3--14.5$\,\mathrm{kpc}$). To describe the full range of the
intensity profile we have added the contribution of the three
components, in other words:
\begin{equation}
I(r)=I(r)_{bulge}+I(r)_{disk}+I(r)_{bar}
\end{equation}
The decomposition of the surface brightness profile was done for
$\it{B, V, R}$ and $\it{I}$ bands. Our best fitted parameters for
each band are given in Table~\ref{tab:parametros} and we show in
Table ~\ref{tab:sersicn} the index {\it{n}} for the bulge and bar
components. The decomposition profile for $\it{R}$ band is shown
in Figure~\ref{fig:descomposicion}.
\begin{table*}
  \caption[]{Galaxy parameters obtained from the profile fits.}
  \label{tab:parametros}
   \begin{center}
    \begin{tabular}{ccccccc}

\hline\hline Filter&$\mu_{e}$&$r_{e}$&$\mu_{0}$&$h$&$\mu_{bar}$&$r_{bar}$ \\
    & mag arcsec$^{-2}$&($\arcsec$)& mag arcsec$^{-2}$&($\arcsec$)&mag arcsec$^{-2}$&($\arcsec$)\\
\hline\noalign{\smallskip}

$B$ & 21.05$\,\pm\,$0.03 & 2.72$\,\pm\,$0.04 &  23.04$\,\pm\,$0.51  & 8.84$\,\pm\,$0.68 & 24.94$\,\pm\,$0.51 & 12.13$\,\pm\,$0.24 \\

            $V$ & 20.19$\,\pm\,$0.02 & 2.79$\,\pm\,$0.03 & 22.28$\,\pm\,$0.54  &

 9.18$\,\pm\,$0.55 & 23.92$\,\pm\,$0.47 & 12.38$\,\pm\,$0.22  \\

            $R$ & 19.88$\,\pm\,$0.02 & 2.95$\,\pm\,$0.03 &  21.10$\,\pm\,$0.09  &

 8.35$\,\pm\,$0.28 & 24.04$\,\pm\,$0.19 & 12.76$\,\pm\,$0.26  \\

           $I$& 19.24$\,\pm\,$0.05 & 2.92$\,\pm\,$0.03 &  20.52$\,\pm\,$0.36  &

 8.35$\,\pm\,$0.82 & 23.09$\,\pm\,$0.60 & 13.52$\,\pm\,$0.45  \\
\hline\noalign{\smallskip}
   \end{tabular}

  \end{center}
\end{table*}
\begin{table*}
  \caption[]{S\'ersic index and bulge--disk luminosity ratio.}
\begin{center}
   \begin{tabular}{cccc}
\hline\hline
 Filter & S\'ersic $n$ & S\'ersic $n$ & $B/D$\\
&(\small{\it{bulge}}) & (\small{\it{bar}}) &\\
\hline\noalign{\smallskip}

 $B$&0.77$\,\pm\,$0.04 & 0.33$\,\pm\,$0.12 & 1.00$\,\pm\,$0.78 \\
 $V$&0.78$\,\pm\,$0.03 & 0.36$\,\pm\,$0.11 & 1.09$\,\pm\,$0.83 \\
 $R$&0.58$\,\pm\,$0.02 & 0.36$\,\pm\,$0.05 & 0.58$\,\pm\,$0.10\\
 $I$&0.66$\,\pm\,$0.05 & 0.33$\,\pm\,$0.20 & 0.63$\,\pm\,$0.37 \\
 \hline\noalign{\smallskip}
\end{tabular}
\label{tab:sersicn}

Note.--- The S\'ersic index applied for the disk component\\ in the model
profile was fixed to {\it{n}}=1, an exponential disk.
\end{center}
\end{table*}

\subsubsection{Bulge and disk parameters}
\label{subsubsec:bulge}

The total luminosity described by a S\'ersic $r^{1/n}$ profile is
given by \citet{G96} as
\begin{equation}
 \label{ec:luminosidad}
L_{tot}=\int_{0}^{\infty} I(r)2\pi rdr=\frac{n2\pi r_{e}^2I_{e}e^{b_{n}}}{b_n^{2n}} \Gamma(2n)
\end{equation}
where $\Gamma$ is the gamma function and $b_n$ is the parameter
defined in previous section. Therefore, the bulge-to disk
luminosity ratio is given by the expression:
\begin{equation}
\frac{B}{D}=\frac{n_be^{b_n}}{b^{2n_b}}\left(\frac{r_e^2}{h^2}\right)\left(\frac{I_e}{I_0}\right)
\Gamma(2n)
\end{equation}
where the $B$ and $D$ are the bulge and disk luminosities of the
galaxy; the index $n_b$ is the S\'ersic index for the bulge;
$r_e$, $I_e$, $h$ and $I_0$ are the parameters related with the
bulge and the disk, respectively.
The integral of intensity shown in equation~\ref{ec:luminosidad}
only has an analytical expression for integer values of
{\it{n}}$\,\geq\,1$ and in some particular cases where
$0\,\leq\,n\leq\,1$ (e.g. 0.2, 0.4). Table~\ref{tab:sersicn} shows
that the bulge component index {\it{n}} varies between 0.77 in {\it{B}}
and 0.58 in {\it{R}} band. For this reason, to estimate $B/D$ we
considered the bulge luminosity separately doing an interpolation
for the corresponding values of {\it{n}}. This interpolation was
obtained from the values where the Equation ~\ref{ec:luminosidad}
has analytical solution. The values obtained for the ratio $B/D$
are shown in Table~\ref{tab:sersicn}.

For a sample of Seyfert galaxies \citet{Chatz01} obtained a wide range of
values for $\mu_{e}$, with average values of 17.64 for Sy 1 and
18.77 for Sy 2, while for $r_e$ reports values of 1.35 and 1.06,
respectively. Also, this author has found that typical values for both
types of Seyfert galaxies are $h\sim\,1-10\,\mathrm{kpc}$ and
$\mu_0\sim\,18-24\,\mathrm{mag\,arcsec}^{-2}$. Thus, our estimated
values of $\mu_e$, $\mu_0$ and $h$ are in agreement with the
values reported by Chatzichristou for Sy galaxies.

Using the classification scheme established for a sample of field
galaxies from HST Medium Deep Survey \citep[]{S97}, we can see
that the obtained value $B/D=0.58$ in the {\it{R}} band for  SBS~0748+499
falls in the  range $0.5<B/D<1.7$, so this object can be
classified as an intermediate type galaxy (SBab). However, the
typical spiral structure it is not clearly present in this object.

\section{Spectral Classification}
\label{sec:espectro}

\begin{figure}[!t]
\includegraphics[width=\columnwidth]{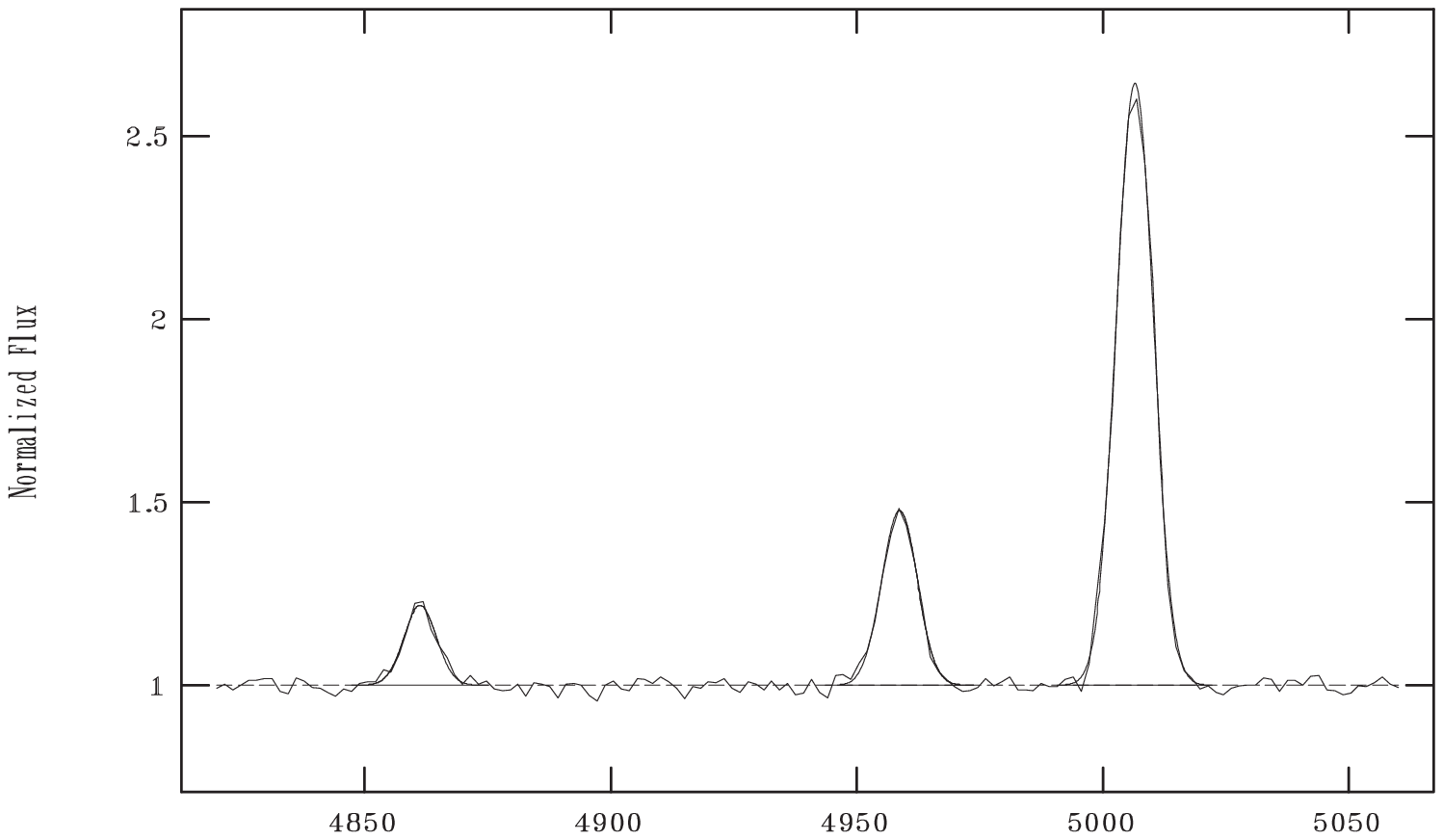}
\includegraphics[width=\columnwidth]{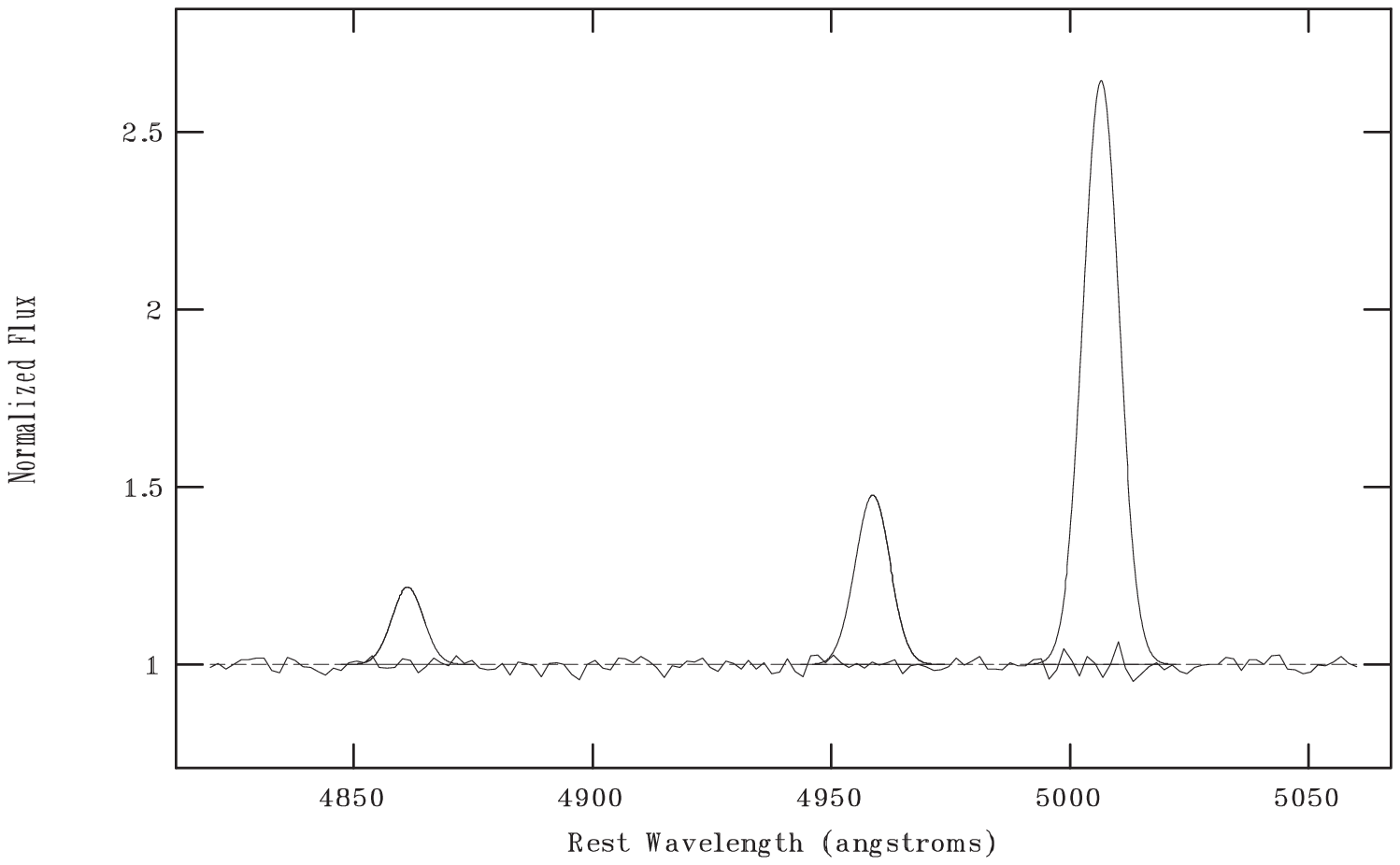}
\caption{Deredshifted smoothed blue normalized spectra of
SBS~0748+499. The upper panel shows the best-fit with three narrow
Gaussian components profiles for H${\beta}$ and [OIII]. The bottom
panel shows individual Gaussian components and the differences
between the data and the best fit, i. e., the residuals }
\label{fig:blue spectra}
\end{figure}

\begin{figure}[!t]
\includegraphics[width=\columnwidth]{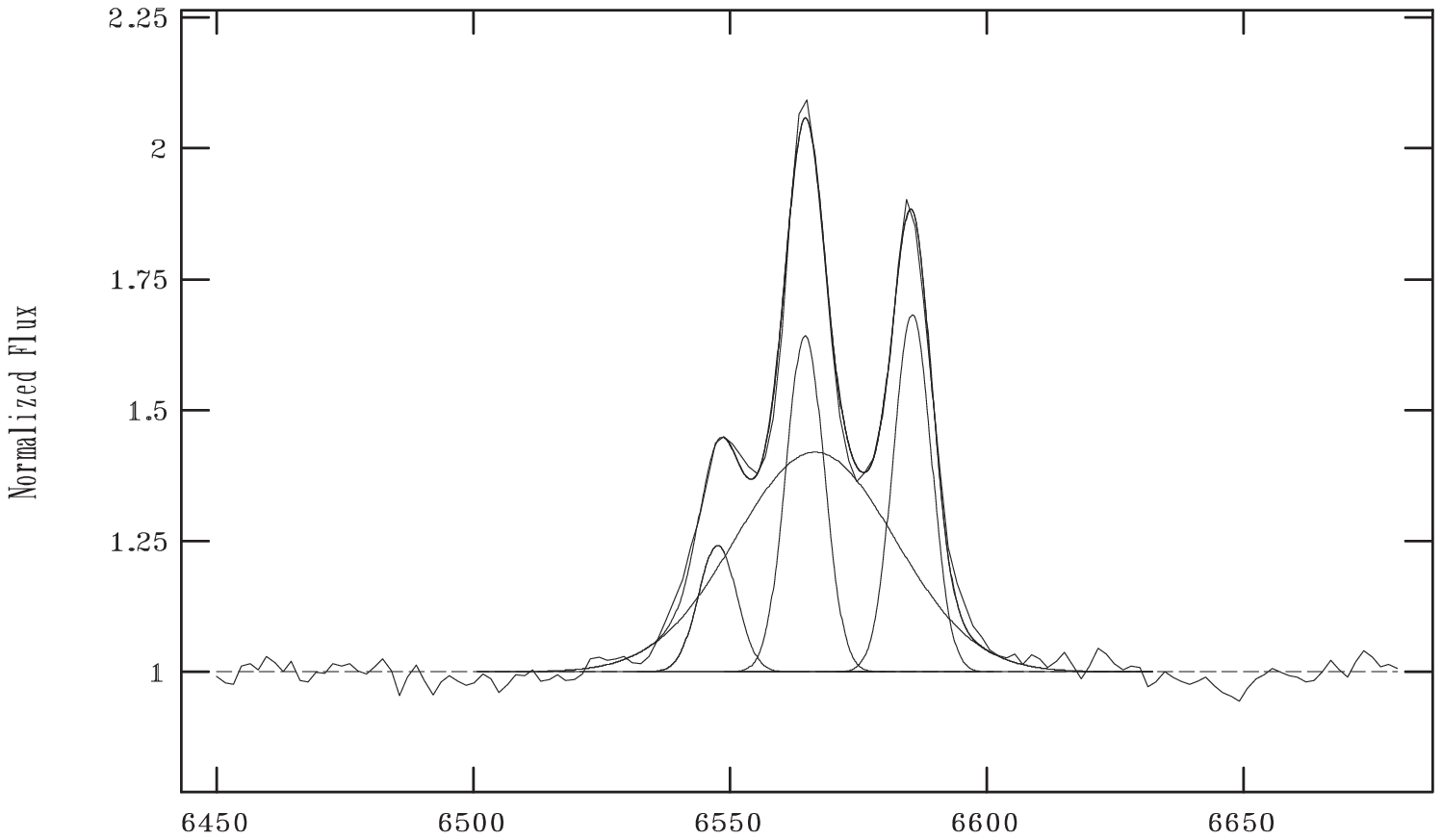}
\includegraphics[width=\columnwidth]{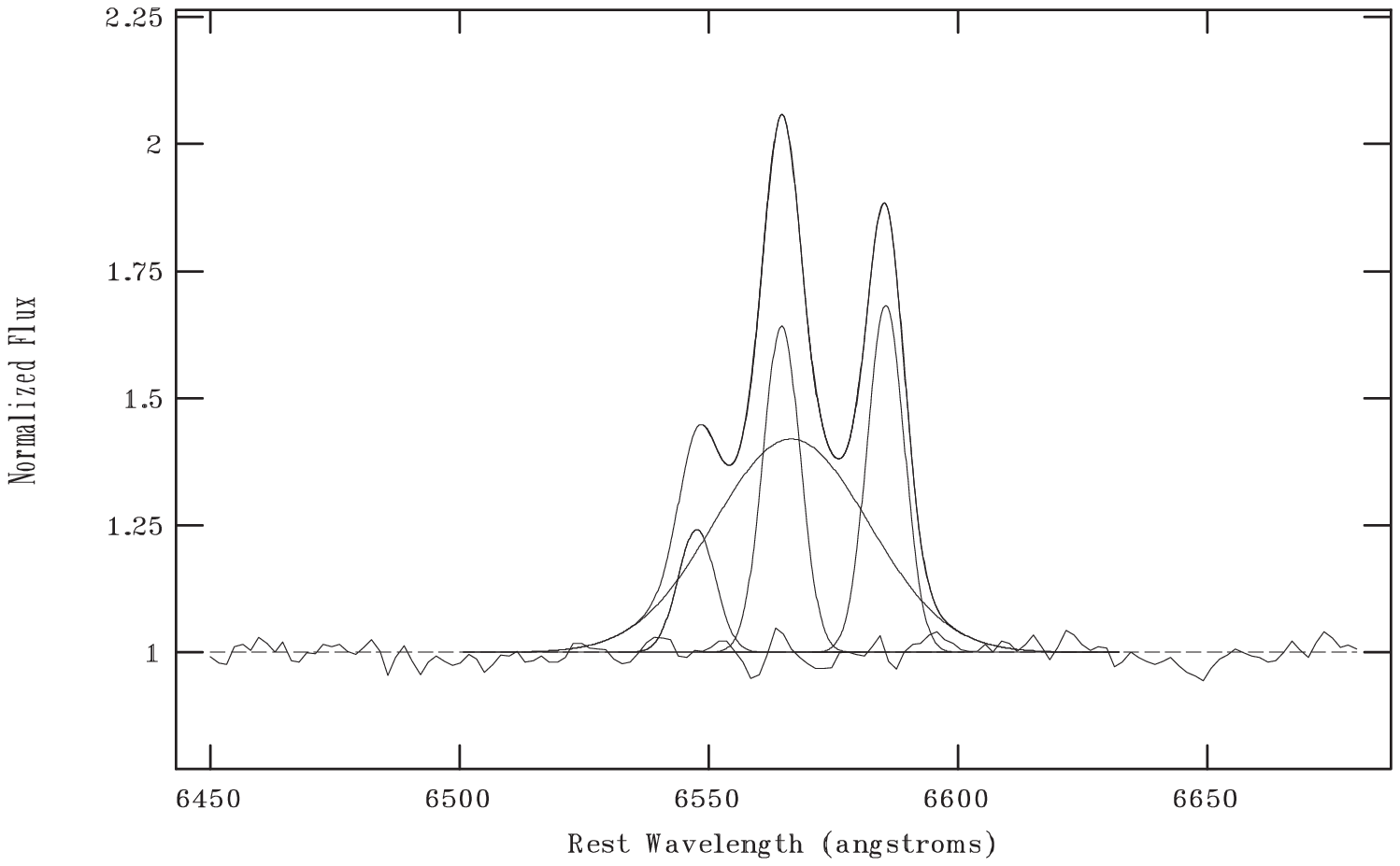}
\caption{Deredshifted smoothed red normalized spectra of
SBS~0748+499. The upper panel shows the best-fit using three
narrow plus one broad Gaussian components profiles for H${\alpha}$
and [NII]. The bottom panel shows individual Gaussian components
and the differences between the data and the best fit.}
\label{fig:red spectra}
\end{figure}

Line fitting analysis for the optical spectra of SBS~0748+499 was done
in terms of Gaussian components \citep[]{Ve81}.  Also, the width and
redshifts of each narrow line component were fixed to the same
value. For the fitting, we assumed that the intensity ratios of [NII]
$\lambda\lambda$ 6548,6583 and [OIII] $\lambda\lambda$ 4959,5007 lines
were equal to 3 and 2.96, respectively \citep[]{Os89}. For H${\alpha}$
and [NII], we have done the fitting using gaussian components (three
narrow and one broad) in the centroid of H$\alpha$. For H${\beta}$ and
[OIII]$\lambda\lambda$ 4959,5007 lines, we used three narrow Gaussian
components for the fitting. The local continuum in the H${\beta}$ and
H${\alpha}$ spectral ranges was approximated inside 4500--5500 \AA\
and 6000--7000 \AA\ regions, respectively. Results are given in
Table~\ref{tab:datos} and in Figures~\ref{fig:blue spectra} and
~\ref{fig:red spectra}. A least-square fitting was performed using the
IRAF/GUIAPPS/SPECTOOL package, and is shown in Figure
~\ref{fig:blue spectra}. In order to classify SBS~0748+499, we used
the diagnostic diagrams $\mathrm{log}$([OI]$\lambda\,$6300/H$\alpha$),
$\mathrm{log}$([NII]$\lambda$\,6583/H$\alpha$), and
$\mathrm{log}$([SII]$\lambda\lambda\,$6717+6731/H$\alpha$) vs.
$\mathrm{log}$([OIII]$\lambda\,$5007/H$\beta$) \citep[]{VO87}.  The
values for logarithms, [OI]$\lambda\,$6300/H$\alpha$=$-$0.54,
[NII]$\lambda\,$6583/H$\alpha$=$-$0.04,
[SII]$\lambda\lambda\,$6717+6731/H$\alpha$=$-$0.44, and
[OIII]$\lambda\,$5007/H$\beta$=$-$0.87 locate the object, on all of
the diagrams, inside the AGN region.

Thus, according to \citet{Os81}, SBS~0748+499 can be classified
as Seyfert 1.9, mainly because we can only see the broad H${\alpha}$
component with a FWHM=1700\,km\,s$^{-1}$. The narrow lines
that appear in Table~\ref{tab:datos} all have a FWHM=450\,km\,s$^{-1}$.


\begin{table*}
  \caption[]{Spectroscopic Data}
   \label{tab:datos}
\begin{center}
    \begin{tabular}{cccc}
\hline\hline Ion & $\lambda$ & Flux & EW\\
            &\small{(\AA)}&\small{($10^{-15}\,\mathrm{erg}\,\mathrm{s}^{-1}\,\mathrm{cm}^{-2}$)}&\small{(\AA)}\\

\hline\noalign{\smallskip}

    $[$\ion{S}{II}$]$ & 6731 & 2.7 & 2.0\\
    $[$\ion{S}{II}$]$ & 6716 & 1.9 & 1.4\\
    $[$\ion{N}{II}$]$ & 6583 & 11.7 & 7.6\\
    H$\alpha$ \small{broad} & 6567 & 22.2 & 14.4\\
    H$\alpha$ \small{narrow} & 6563 & 12.8 & 8.3\\
    $[$\ion{N}{II}$]$ & 6548 & 3.9 & 2.5\\
    $[$\ion{O}{I}$]$ & 6300 & 3.7 & 2.5\\
    $[$\ion{O}{III}$]$ & 5007 & 23.7 & 14.0\\
    $[$\ion{O}{III}$]$ & 4959 & 8.0 & 4.6\\
    H$\beta$ & 4861 & 3.2 & 1.9\\

\hline\noalign{\smallskip}
   \end{tabular}

Note.---All parameters are given in the rest frame.
\end{center}
\end{table*}

\section{Black Hole Mass Estimation}
\label{sec:BH}

We have used two different methods to give an estimation of the
black hole mass of SBS 0748+499. The first one is based on the
absolute $R$-band magnitude of the bulge component obtained from
the photometric data, i.e., from the host galaxy. The second one is
based on an estimation of the stellar velocity dispersion of the
bulge which was derived indirectly from our spectroscopic data.
The first method is based on the application of the relation
between absolute $R$-band magnitude $M_R$ and $M_{BH}$ given by
\citet{MD02}. We have recalculated this relation for our cosmology
as:

\begin{equation}
\log(M_{BH}/M_{\odot})=-0.50(\pm\,0.02)M_R-2.52\,(\pm\,0.48)
\end{equation}
From the luminosity of the bulge component in $R$--band
$L_R=(5.17\,\pm\,0.21)\,\times\,10^9\,L_{\odot}$ we have
$M_R=-19.86\,\pm\,0.05$, and we obtain
$\mathrm{log}\,M_{BH}/M_{\odot}=7.41\,\pm\,0.48$.

The second method needs an indirect inference of the bulge stellar
velocity dispersion. The brightness of the nuclear part of SBS
0748+499 does not allow us to detect the absorption lines of CaT
in the spectrum (lines commonly used to estimate $\sigma_{\star}$,
e.g.  \citet{On04}). Therefore, we decide to use the
relation found by \citet{N00}: $\sigma_{\star}=$\,FWHM
[OIII]$_{5007}$/2.35, on the assumption that for this galaxy, the
forbidden line kinematics is dominated by virial motions in the
host galaxy bulge. Also, this author found the following
relation for the black hole mass:

\begin{equation}
 \log(M_{BH}/M_{\odot})=3.7\,(\pm0.7)\log\sigma_{\star}-0.5\,(\pm\,0.1)
\end{equation}

This method has been investigated later by \citet{B03} using the Sloan
Digital Sky Survey Early Data Release spectra of 107 low redshift
quasars and Sy1 galaxies. He confirms the existence of the relation
$\sigma_{\star}$ versus FWHM [OIII]$_{5007}$ and claims that this
relation can predict the black hole mass in active galaxies to within
a factor of 5. So, measuring the FWHM of the narrow line
[OIII]$_{5007}$\,=\,$450\,\pm\,14$\,km\,s$^{-1}$ from our spectrum
(see $\S$~\ref{sec:espectro}) we obtain
$\sigma_{\star}=(191\,\pm\,6)\mathrm{km\,s^{-1}}$ which yields a value
for the mass of the black hole of
$\log\,M_{BH}/M_{\odot}=7.94\,\pm\,1.6 $.

The values found with the methods described above differ by a
factor of 3, and the photometric method developed by \citet{MD02}
provides an estimation with lower dispersion.

\section{Discussion and Conclusions}

\label{sec:discusion}

From the $B$-band image in Figure~\ref{fig:RGB} and also from the
surface brightness profiles (see Figure~\ref{fig:brillo}) we can
clearly distinguish a bulge and a disk component, as well as a
bar--like structure in the host galaxy of SBS~0748+499.  This is
important since bar-induced inflows are frequently cited mechanisms
for funnelling some gas to the nucleus, thereby temporarily activating
an otherwise normal galaxy. Additionally, from the geometric profiles
we infer the presence of low-surface brightness spiral structure. In
\S~\ref{subsubsec:bulge} we obtained a value of 0.58 for the
bulge--to--disk luminosity ratio in $R$ band, which suggests that
SBS~0748+499 is an SBab type. We can barely appreciate the spiral-arms
structure in Figure~\ref{fig:RGB}, however we could not distinguish
any star forming regions, typically expected in Sab type galaxies. On
the other hand, we have done a comparison between the color $B-V=0.84$
obtained in \S~\ref{sec:mag} and the mean value reported for galaxies
S0a, Sa $B-V=0.78$ by \citet{RH94}. We found that SBS~0748+499 is only
$0.06\,\mathrm{mag}$ redder than these early--type galaxies. This
could indicate that SBS~0748+499 is earlier than a Sb
galaxy. \citet{Gro02} found that Sy2 and LINERS are preferentially
found in redder hosts $B-V=0.92$ while Sy 1 hosts typically have bluer
colors $B-V=0.70$. We have shown in section \S~\ref{sec:espectro} that
SBS~0748+499 is a Sy 1.9, so the color range found for it will fill
the gap between Sy1 and Sy2, supporting that it is an intermediate Sy
type galaxy.  Additionally, we have estimated that the optical
luminosity of SBS~0748+499 is $L_{B}=6.6\times10^{42}\,\mathrm{erg}\,
\mathrm{s}^{-1}$ ($M_{B}=-19.9$), a typical value for Seyfert galaxies
\citet{SG83}.

From its optical spectrum, it is clear that SBS~0748+499 does not
shows the H$\beta$ broad component, therefore the only way to estimate
$M_{BH}$ was using the two different correlations described in the
last section. Our main hypothesis for using the FWHM [OIII]$_{5007}$
(i.e. the NLR in intermediate Seyferts) is based on the fact that this
quantity should be related to the stellar velocity dispersion of the
host galaxy, and therefore with the $M_{BH}$ as it has been previously
found. On the other hand, the $M_R$ bulge in this work was
obtained using the best fitted parameters excluding the innermost
2\arcsec region.

The relation that involves $M_R$ bulge provides an
estimation of the $M_{BH}$ that is within the uncertainties
obtained with the FWHM [OIII]$_{5007}$-- $\sigma_{\star}$
relation. At this point, we consider that both estimates are in
good agreement, and that the photometric method provides lower
uncertainties. Also, it is interesting to note that both
correlations are calibrated and based on Type 1 AGN (Sy1 and QSO),
and also that \citet{MD02} have demonstrated that the $M_R$
bulge -- $M_{BH}$ correlation is also valid for quiescent galaxies.

SBS~0748+499 is a nearby AGN, so it can be studied in more detail.  In
future work, it would be worthy to study a well defined sample of
nearby intermediate Seyfert galaxies in order to estimate the
$\sigma_{\star}$ and see if in particular, the so--called
$M_{BH}$-$\sigma_{\star}$ scales in the same form as it does for Sy1
galaxies. In particular, new observations with better resolution are
needed to establish the nuclear morphology in the innermost arcsecond
region of SBS~0748+499.

\acknowledgements

The authors want to acknowledge our anonymous referee for very
useful comments and suggestions, which helped to improve this
work, and to Dr. Lee who kindly corrected the English version. J.
T. acknowledges financial support from grant IN-118601
(PAPIIT-UNAM) scholarship. E. B. acknowledges grant IN-112305
(PAPIIT-UNAM) for the support given to this research. V. Ch.
acknowledges CONACyT basic research grant No.39560-F. We also
acknowledge the OAN-SPM and OAGH staff for their support during
the observations. This research has made use of the NASA/IPAC
Extragalactic Database (NED) which is operated by the Jet
Propulsion Laboratory, California Institute of Technology, under
contract with the National Aeronautics and Space Administration.

\end{document}